\documentclass[10pt]{article}
\usepackage{cite}
\usepackage{graphicx,amsmath,amssymb,amsbsy,latexsym,amsthm}
\usepackage{bm}
\usepackage[mathscr]{eucal}

\theoremstyle{plain}
\newtheorem{definition}{Definition}
\newtheorem{theorem}{Theorem}

\newtheorem{lemma}{Lemma}
\newcommand{\startproof}{\noindent\textbf{Proof.} }
\newcommand{\finishproof}{\hfill $\blacksquare$ \\}

\numberwithin{equation}{section}

\def\R{\mathbb{R}}

\def\C{\mathbb{C}}

\def\tr{\mathrm{tr}}
\def\half{\frac{1}{2}}

\newcommand{\so}{\mathfrak{so}}

\newcommand{\su}{\mathfrak{su}}

\newcommand{\spin}{\mathfrak{spin}}
\newcommand{\mskew}{\mathrm{skew}}

\newcommand{\eqa}{\begin{eqnarray}}
\newcommand{\neqa}{\end{eqnarray}}
\newcommand{\be}{\begin{equation}}
\newcommand{\ee}{\end{equation}}

\newcommand{\scr}[1]{\mathcal{#1}}

\newcommand{\dual}{\,\,{}^\star\!}

\newcommand{\dif}{\mathrm{d}}

\newcommand{\Hil}{\mathcal{H}}

\newcommand{\geom}{\mathrm{geom}}
\newcommand{\phys}{\mathrm{phys}}
\newcommand{\Regge}{\mathrm{Regge}}

\newcommand{\sut}[1]{{\bm #1}}
\newcommand{\cldataSUtwo}{\dummy reduced boundary data \dummy}

\newcommand{\qdata}{\dummy quantum \dummy}
\newcommand{\Qdata}{\dummy Quantum \dummy}

\newcommand{\spinsp}{V}
\newcommand{\rotk}{k}  
\newcommand{\simpj}{s}
\newcommand{\genj}{j}
\newcommand{\genjpm}{j}
\newcommand{\lcd}{\eta}
\newcommand{\norm}{n}

\newcommand{\unitpm}{u}
\newcommand{\simp}{S}
\newcommand{\geosimp}{\sigma}
\newcommand{\geoname}{\dummy geometrical 4-simplex \dummy}
\newcommand{\Geoname}{Geometrical 4-simplex}
\newcommand{\ordname}{\dummy ordered 4-simplex \dummy}
\newcommand{\Ordname}{Ordered 4-simplex}
\newcommand{\anordname}{\dummy an ordered 4-simplex \dummy}

\newcommand{\ggk}{Y}
\newcommand{\rotgen}{L}
\newcommand{\spinset}{\mathcal{K}_\gamma}
\newcommand{\dummy}{\rule{0mm}{0mm}}

\begin{document}

\title{The Plebanski sectors of the EPRL vertex}

\author{Jonathan Engle\thanks{jonathan.engle@fau.edu}
 \\[1mm]
\normalsize \em Department of Physics, Florida Atlantic University, Boca Raton, Florida, USA}
\date{\today}
\maketitle\vspace{-7mm}

\begin{abstract}

Modern spin-foam models of four dimensional gravity are
based on a discrete version of the $Spin(4)$ Plebanski formulation.
Beyond what is already in the
literature, we clarify the meaning of different Plebanski sectors in this classical discrete model.  We show that the linearized simplicity constraints used in the EPRL and FK models
are not sufficient to impose a restriction to a single Plebanski sector, but rather, three Plebanski sectors are mixed.  We propose this as the reason for certain
extra `undesired' terms in the asymptotics of the EPRL vertex analyzed
by Barrett et al. This explanation for the extra terms is new and different from that sometimes offered  in the spin-foam literature thus far.
\end{abstract}


\section{Introduction}

Spin-foams \cite{rovelli2004, perez2003, rovelli2011} provide a path integral framework for defining the dynamics of
loop quantum gravity (LQG)\cite{al2004, rovelli2004, thiemann2007, rovelli2011},
a background independent, canonical quantization of general relativity.  Ever since the Barrett-Crane model \cite{bc1997},
spin-foams have been based on the \textit{$Spin(4)$ Plebanski} formulation of gravity \cite{df1998, plebanski1977},
in which the \textit{co-tetrad} $e$
is replaced by an $\mathfrak{so}(4)$-valued 2-form $B = \dual e \wedge e$,
where $\dual$ denotes Hodge duality in the internal tetrad indices; at the same time, one introduces a constraint, called the simplicity constraint,
to ensure $B$ is of this form for some $e$.  However, the original constraint \textit{function} used in
\cite{df1998, bc1997, bhnr2004} (equation (\ref{plebconstr}))
does \textit{not} impose $B = \dual e \wedge e$ for some $e$. Rather, it imposes
that $B$ belong to one of five possible sectors:
\begin{enumerate}
\item[(I$\pm$)] $B = \pm e \wedge e$ for some $e$
\item[(II$\pm$)] $B = \pm \dual e \wedge e$ for some $e$
\item[(deg)] $B$ is degenerate
($\tr(\dual B \wedge B) = 0$)
\end{enumerate}
where the terminology for the sectors has been taken from \cite{bhnr2004}.  We call
these \textit{Plebanski sectors}.

The only spin-foam model thus far to match the kinematics of LQG is the EPRL model \cite{elpr2007, rovelli2011}.
This, together with the FK model \cite{fk2007}, form the so-called ``new models''.  The common feature of both of these
models is a certain reformulation of the simplicity constraints that is
\textit{linear} in the $B$ variables.
%
%
Existing literature does not directly address the question of which of the
Plebanski sectors are still included by the linear constraints.
In this paper we answer this question, and show in a precise sense
that sectors (II$\pm$) and (deg) are included.
We furthermore show that this mixing of Plebanski sectors
is precisely the reason for the ``undesired'' terms in the semiclassical limit of the EPRL
vertex computed in \cite{bdfgh2009}. This explanation is new and different from the explanation in terms of a sum over
orientations of simplices mentioned, for example, in \cite{mp2011}.  We discuss this
issue in the conclusion section. As section \ref{clsect} is
classical in nature and is an analysis of the linear simplicity constraints which the FK model also uses, all discussion and
conclusions in section \ref{clsect}
are equally relevant for FK as for EPRL.

The paper is structured as follows.
We begin by reviewing the classical discrete framework underlying the EPRL and FK models.  We then clarify the meaning
of Plebanski sectors in this discrete context, and prove that the linear simplicity constraint imposes
a restriction to precisely Plebanski sectors (II$\pm$) and (deg).
Finally, we review the asymptotics of the EPRL vertex, and identify the
Plebanski sector of the critical point of each term.
We then close with a discussion.

%
%

\section{Classical analysis of the linear simplicity constraints}
\label{clsect}

\subsection{Discrete classical framework}
\label{clframe}

\subsubsection{Generalities}

First, let us introduce some conventions and definitions.
Elements of $\su(2)$ will be bold.
We use the normalized basis $\sut{\tau}^i := \frac{-i}{2}\sut{\sigma}^i$ of $\su(2)$,
where $\sut{\sigma}^i$ are the standard Pauli matrices, so that $[\sut{\tau}^i, \sut{\tau}^j]= \epsilon^{ij}{}_k \sut{\tau}^k$.
Given an element $\sut{\lambda} \in \su(2)$, $\lambda^i \in \R^3$
shall denote its components with respect to $\sut{\tau}^i$.
Let $I$ denote the $2\times 2$ identity matrix.
%
%
In the following, we will freely use the natural
isomorphism between $\so(4)$ and $\spin(4):= \su(2) \oplus \su(2)$,
$J^{IJ} \leftrightarrow (\sut{J}_-,\sut{J}_+) \equiv (J_-^i\sut{\tau}_i, J_+^i\sut{\tau}_i)$,
effectively identifying these two algebras. Here $J^{IJ} = - J^{JI}$ and
$I,J  = 0,1,2,3$.
Explicitly, the isomorphism is
\begin{equation}
J_{\pm}^i = \frac{1}{4}\epsilon^i{}_{jk}J^{jk} \pm \frac{1}{2}J^{0i}.
\end{equation}
with inverse
\begin{eqnarray}
\nonumber
J^{ij} &=& \epsilon^{ij}{}_k (J_+^k + J_-^k) \\
\label{alginv}
J^{0i} &=& J_+^i - J_-^i .
\end{eqnarray}
%
%

\subsubsection{Continuum theory}

The EPRL model, as all spin-foam models of gravity, takes as its starting point the formulation
of gravity as a \textit{constrained BF theory}, following the ideas of
Plebanski \cite{plebanski1977}.  The basic continuum variables are an $\so(4)$-valued 2-form
$B_{\mu\nu}^{IJ}$ and an $\so(4)$ connection $\omega_\mu^{IJ}$.
We call $B_{\mu\nu}^{IJ}$ the \textit{Plebanski two-form}.
The action is
\begin{equation}
\label{BFaction}
S = \frac{-1}{2\kappa}\int \tr[(B + \frac{1}{\gamma} \dual B) \wedge F] .
\end{equation}
where $F := \dif \omega + \omega \wedge \omega$ is the curvature of $\omega$, and $\kappa := 8\pi G$.
The variable conjugate to $\omega$ is thus
\begin{displaymath}
J := \frac{1}{\kappa} (B + \frac{1}{\gamma} \dual B) .
\end{displaymath}
%
%
In terms of the anti-self-dual and self dual parts of $J$ and $B$,
\begin{equation}
\label{bjsd}
(J^\pm)^i = \left(\frac{\gamma \pm 1}{\kappa\gamma}\right) (B^{\pm})^i
\end{equation}
%
%
The action (\ref{BFaction}) is a \textit{BF action}, and leads to a topological field theory
with no local degrees of freedom.  To turn it into gravity, one  imposes the
\textit{simplicity constraints}, reviewed here in equations (\ref{plebconstr}) and
(\ref{discsimp}).

\subsubsection{Discrete variables}
\label{discvarsect}

To construct the spin-foam model, one introduces a discretization of
space-time using a triangulation into 4-simplices. But for the purpose of analyzing the
vertex amplitude, it suffices to focus on a single 4-simplex.  We therefore do so.
In the past years \cite{kkl2009,dhr2010}
the EPRL model has also been generalized to arbitrary cell-complexes; however, for
simplicity, and because we heavily use the work \cite{bdfgh2009},
we restrict ourselves to the case in which the cell-complex
is a simplicial triangulation.

Consider an oriented 4-simplex $\simp$. Number the tetrahedra $a=0, \dots ,4$,
%
%
and label the triangles by
the unordered pair $(ab)$ of tetrahedra that contain it. One
thinks of each tetrahedron, as well as the 4-simplex itself, as having its own `frame' \cite{epr2007}.
%
%
The connection $\omega$ is discretized
by specifying a parallel transport map from each tetrahedron to the 4-simplex
frame -- thus,
in our case, there are 5 parallel transport maps $G_a = (X_a^-,X_a^+)$, $a = 1, \dots ,5$. Let $\Delta_{ab}$ denote
the triangle $(ab)$, endowed with the orientation induced on it as part of the
boundary of tetrahedron $a$, whose orientation in turn is induced from its being part of the boundary
of $\simp$.
The two-forms $B$ and $J$ are then discretized as the elements
\begin{displaymath}
B_{ab} = \int_{\Delta_{ab}} B, \quad J_{ab} = \int_{\Delta_{ab}} J
\end{displaymath}
where one thinks of these elements as being `in the frame at $a$.'
For each $ab$, these algebra elements are related, in terms of their self-dual and anti-self-dual parts, by
\begin{equation}
\label{sdasd_rel}
(J^\pm_{ab})^i = \left(\frac{\gamma \pm 1}{\kappa\gamma}\right) (B^{\pm}_{ab})^i.
\end{equation}
Because the bivectors $J_{ab}$ `in the frame at $a$' are key in the 
canonical theory in section \ref{EPRLsec}, we call them the \textit{canonical bivectors}.

However, in the rest of section \ref{clsect}, we will focus on 
reconstructing the \textit{four-dimensional continuum Plebanksi two-form}
$B_{\mu\nu}^{IJ}$ from this discrete data, 
and proving related results.
%
%
For this purpose,  it is necessary to parallel transport all of the
bivectors $B_{ab}^{IJ}$ to a \textit{common} frame.  
If one parallel transports them to the 4-simplex frame,
one obtains
\begin{displaymath}
B_{ab}(\simp) := G_a \triangleright B_{ab}, 
\end{displaymath}
where $\triangleright$ denotes the adjoint action of $Spin(4)$ on $\spin(4) \cong \so(4)$.
If one parallel transports them to some 
tetrahedron frame $c$, one has
\begin{displaymath}
B_{ab}(t_c) := G_{ca} \triangleright B_{ab}.
\end{displaymath}
where $G_{ca}:= G_c^{-1} G_a$ is the parallel transport map from tetrahedron
$a$ to tetrahedron $c$.
Throughout the rest of this section, 
in order to use a presentation closer to that in \cite{bdfgh2009},
%
%
we will work in the 
4-vertex frame; however, all constructions and analyses in the rest of this section
can be equally done in any of the tetrahedron frames --- 
when a tetrahedron frame is used, only
the parallel transports $G_{ab} := G_a^{-1} G_b$ between tetrahedra are
needed.

Let us now turn to some expected properties of the bivectors $B_{ab}$.
Note that in the 4-simplex frame, because $\Delta_{ab}, \Delta_{ba}$ differ
by only a change of orientation, one should have
\begin{equation}
B_{ab}(\simp) \equiv - B_{ab}(\simp).
\end{equation}
This is the `orientation constraint'. As was shown in \cite{bdfgh2009},
this is imposed by the EPRL
vertex amplitude itself: If it is not satisfied, the vertex amplitude is exponentially
suppressed.
Furthermore, classically, in each tetrahedron frame $a$, the four algebra elements $B_{ab}$ must satisfy the
\textit{closure relation} $\sum_{b \neq a} B_{ab} = 0$. This is similarly imposed by the vertex amplitude being
exponentially suppressed if it is not satisfied.  Finally, if the closure
relation is satisfied in each tetrahedron, then the algebra elements for each tetrahedron uniquely
determine a tetrahedron geometry, and hence a \textit{shape} for each triangle $\Delta_{ab}$ in the tetrahedron.
One then would expect that the shape of the triangle $\Delta_{ab}$ as determined in tetrahedra $a$ and $b$ would
be the same --- the \textit{gluing constraint} \cite{ds2008, dr2008}.
As we shall see, when this is not satisfied, either the vertex amplitude is exponentially
suppressed or one is in the degenerate Plebanski sector.

A constraint that holds from the start in both the classical and quantum
frameworks is $|\sut{B}_{ab}^\pm|^2 = |\sut{B}_{ba}^\pm|^2$.
It is consequently convenient to introduce separate variables for the \textit{norms} and \textit{directions}
of the self-dual and anti-self-dual parts of these bivectors.  In
the tetrahedron frames, for each pair $(ab)$, we then have a pair of norms
$B_{ab}^\pm \equiv B_{ba}^\pm := |\sut{B}_{ab}^\pm|$
and two pair of directions (equivalently, unit $\su(2)$ elements)
$\sut{\unitpm}^\pm_{ab} := \sut{B}^\pm_{ab}/|B^\pm_{ab}(\simp)|$, $\sut{\unitpm}^\pm_{ba} := \sut{B}^\pm_{ba}/|B^\pm_{ba}(\simp)|$.  One has
\begin{displaymath}
B_{ab} = (B_{ab}^- \sut{\unitpm}_{ab}^-, B_{ab}^+ \sut{\unitpm}_{ab}^+).
\end{displaymath}
From (\ref{sdasd_rel}), also
\begin{equation}
%
%
J_{ab} = \left( \left(\frac{\gamma - 1}{\kappa \gamma}\right)B_{ab}^- \sut{\unitpm}_{ab}^-,
\left(\frac{\gamma + 1}{\kappa \gamma}\right)B_{ab}^+ \sut{\unitpm}_{ab}^+\right) .
\end{equation}
The bivectors in the simplex frame are then given by
\begin{equation}
\label{Bsimp_param}
B_{ab}(\simp) = (B_{ab}^- \sut{\unitpm}_{ab}^-(\simp), B_{ab}^+ \sut{\unitpm}_{ab}^+(\simp)),
\end{equation}
where $\sut{\unitpm}_{ab}^\pm(\simp) := X_{a}^\pm \triangleright \sut{\unitpm}_{ab}^\pm$.
The orientation constraint then
takes the form $\sut{\unitpm}^\pm_{ab}(\simp) = - \sut{\unitpm}^\pm_{ba}(\simp)$.

\subsection{Discrete Plebanski sectors defined}

We now define what we call a ``discrete Plebanski field''.
As will be seen in definition \ref{bivgeomdef}, it plays the role of a precursor
to the bivector geometry definition used in \cite{bdfgh2009}.
\begin{definition}[Discrete Plebanski field]
A discrete Plebanski field is a set of bivectors $\{B_{ab}^{IJ}\}$, $a,b = 0,1,2,3,4$,
such that
\begin{enumerate}
\item[(i.)] $B_{ab}^{IJ} = - B_{ba}^{IJ}$ (orientation)
\item[(ii.)] $\sum_{b \neq a} B_{ab}^{IJ} = 0$ (closure)
\end{enumerate}
\end{definition}
%
%
In the following we will see that every discrete Plebanski field determines
in a certain sense a \textit{continuum} $\so(4)$-valued two-form
field in the simplex. This will then be used to define the notion of a discrete
Plebanski field being in a certain ``Plebanski sector''.

Let $M$ denote $\R^4$ as an oriented manifold, equipt with the
canonical flat connection $\partial_a$ on $\R^4$.
This is the arena where we will define the simplex and reconstruct the
continuum $\so(4)$-valued two-form from given discrete data
$\{B_{ab}^{IJ}\}$.
The symmetry group of $(M,\partial_a)$ is the proper
inhomogeneous $GL(4)$ group, $IGL(4)^+$.
%
%
$\partial_a$ defines
notions of straight line segments and planes in $M$ in the usual way, and the notion of convex hull is defined
in the usual way using straight line segments.

Furthermore,  let $V$ denote the tangent space of $M$ at
any point.  Because $\partial_a$ is flat,  its parallel transport maps provide
(1.) a natural isomorphism between $V$ and every other tangent space, allowing us
to identify every tangent space with $V$,
and (2.) a natural isomorphism between $V$
and the space of \textit{constant vector fields} on $M$, obtained by parallel transporting any given element of $V$ throughout $M$.  We will let the same symbol denote
a given vector in $V$ and the corresponding vector at any other point, as well
as the corresponding constant vector field on $M$.  In like manner,
$\partial_a$ defines a natural isomorphism between tensors of a given type
over $V$ and tensors of the same type at any other point, as well as with the space
of constant tensors on $M$ of the same type.  Again, We will let the same symbol denote
a given tensor over $V$ and the corresponding tensor at any other point, as well as
the associated constant tensor.

\begin{definition}[\Geoname]
A \geoname in $M$ is the convex hull of 5 points, called vertices,
in $M$, not all of which lie in the same 3-plane.
\end{definition}
\begin{definition}[\Ordname]
\label{ordsimpdef}
We define an \textit{\ordname} $\geosimp$ to be
a \geoname in $M$ with an assignment of
labels $0,1,2,3,4$ to its five vertices such that the ordered set of
four vectors $(\vec{01}, \vec{02}, \vec{03}, \vec{04})$
in $V$ has positive orientation.
Each tetrahedron is then labeled by the number of the one vertex it does not contain.
\end{definition}

A standard way to specify orientation of a manifold is to specify a no-where vanishing
volume form modulo rescaling by a positive function. However, through
the one-to-one relation between volume forms and inverse volume forms,
$\epsilon^{\alpha_1 \cdots \alpha_n} \epsilon_{\alpha_1 \cdots \alpha_n} = n!$,
it is just as easy to specify orientation by a no-where vanishing
\textit{inverse} volume form modulo rescaling by a positive function.
Because it is more convenient,
throughout this section we will specify orientation
in the latter way.
%

\begin{definition}[Oriented triangle $\Delta_{ab}$]
Given \anordname $\geosimp$, let
$\Delta_{ab}=\Delta_{ab}(\geosimp)$ denote the triangle
between tetrahedra $a$ and $b$.
Let $(N_a)_\alpha, (N_b)_\alpha$ be any outward pointing normals
to tetrahedra $a$ and $b$, respectively, and
let $\epsilon^{\alpha\beta\gamma\delta}$ be any
oriented inverse volume form.
Then the inverse 2-form
\begin{equation}
\label{triorient}
\epsilon^{\alpha\beta}_{[ab]} := \epsilon^{\gamma\delta\alpha\beta} (N_a)_{\gamma} (N_b)_{\delta}
\end{equation}
is non-zero and tangent to $\Delta_{ab}$, and well-defined upto rescaling by a positive function.
We let $\epsilon^{\alpha\beta}_{[ab]}$ define the orientation of $\Delta_{ab}$ .
\end{definition}

The above-defined orientation of $\Delta_{ab}$ is simply the orientation induced on $\Delta_{ab}$ as part
of the boundary of tetrahedron $a$, considered as part of the boundary of $\sigma$.\footnote{This is due to the fact that the pull-back of $(N_b)_\alpha$ to the plane of tetrahedron $a$ is an outward pointing normal
to $\Delta_{ab}$.}

\begin{lemma}
\label{twoformex}
Given a discrete Plebanski field $\{B_{ab}^{IJ}\}$ and any choice of
\ordname $\geosimp$ in $M$, there exists a \textit{unique}
constant Lie algebra-valued two-form $B_{\mu\nu}^{IJ}$ such that
\begin{displaymath}
B_{ab}^{IJ} = \int_{\Delta{ab}} B^{IJ}.
\end{displaymath}
\end{lemma}
{\startproof
For each $\Delta_{ab}$ in $\geosimp$, define the bivector $I^{\mu\nu}_{ab}$
by
\begin{displaymath}
I^{\mu\nu}_{ab} \lambda_{\mu\nu} := \int_{\Delta_{ab}} \lambda
\end{displaymath}
for all constant $\lambda_{\mu\nu}$.  This set of bivectors satisfies
the same closure and orientation conditions as $B_{ab}^{IJ}$.
In addition, $\{I^{\mu\nu}_{ab}\}$ satisfies a
non-degeneracy condition: for $i<j$, $i,j = 0,1,2,3$,
$I^{\mu\nu}_{ij}$ form a basis of $V \otimes_{\mskew} V$.
Let $I_{\mu\nu}^{ij}$ denote the corresponding dual basis of
$V^* \otimes_\mskew V^*$.
Then for each $I,J = 0,1,2,3$, define
\begin{equation}
\label{twoformdefeq}
B^{IJ}_{\mu\nu} := \sum_{\substack{i<j \\ i,j = 0,1,2,3}}
B^{IJ}_{ij} I^{ij}_{IJ}.
\end{equation}
We then have
\begin{displaymath}
\int_{\Delta_{ij}} B^{IJ} = I_{ij}^{\mu\nu} B^{IJ}_{\mu\nu} = B^{IJ}_{ij}
\end{displaymath}
for all $i<j$, $i,j = 0,1,2,3$.  Using that both
$B^{IJ}_{ab}, I^{\mu\nu}_{ab}$ satisfy closure and orientation, it follows
\begin{equation}
\label{desiredeq}
\int_{\Delta_{ab}} B^{IJ} = I_{ab}^{\mu\nu} B^{IJ}_{\mu\nu} = B^{IJ}_{ab}
\end{equation}
for all $a,b$, proving existence.
To prove uniqueness, suppose (\ref{desiredeq}) is satisfied.
Then, from the fact that it holds in particular for $a<b, a,b = 0,1,2,3$,
it follows immediately that $B^{IJ}_{\mu\nu}$ is as in
(\ref{twoformdefeq}).
%
%
\finishproof}

\begin{definition}
Given a discrete Plebanski  field $\{B_{ab}^{IJ}\}$ and any choice of
\ordname $\geosimp$ in $M$, we call
the unique $B_{\mu\nu}^{IJ}$ in lemma \ref{twoformex} the
\textit{two-form of $\{B_{ab}^{IJ}\}$ adapted to $\geosimp$}.
\end{definition}

If $B_{\mu\nu}^{IJ}$ satisfies the Plebanski constraint,
\begin{equation}
\label{plebconstr}
\epsilon_{IJKL}(B_{\mu\nu}^{IJ} B_{\rho\sigma}^{KL}
- \frac{1}{4!}\lcd_{\mu\nu\rho\sigma} \lcd^{\alpha\beta\gamma\delta}
B_{\alpha\beta}^{IJ} B_{\gamma\delta}^{KL}) = 0,
\end{equation}
where $\lcd_{\mu\nu\rho\sigma}$ and $\lcd^{\alpha\beta\gamma\delta}$
denote the Levi-Civita tensors of density weight $-1$ and $1$, respectively,
then, for example from \cite{bhnr2004}, it must be of one of the
five forms
\begin{description}
\item[(I$\pm$)] $B^{IJ} = \pm e^I \wedge e^J$ for some $e^I_\mu$
\item[(II$\pm$)] $B^{IJ} = \pm \frac{1}{2} \epsilon^{IJ}{}_{KL} e^K \wedge e^L$
for some $e^I_\mu$
\item[(deg)] $\epsilon_{IJKL}\lcd^{\mu\nu\rho\sigma} B^{IJ}_{\mu\nu} B^{KL}_{\rho\sigma} = 0$ (degenerate case).
\end{description}
Each of these forms defines
a particular sector, which we refer to as (I$\pm$), (II$\pm$), and (deg).
These five sectors are disjoint \cite{bhnr2004}.

\begin{definition}
If, for a given choice of \ordname $\geosimp$,
a given discrete Plebanski field $\{B_{ab}^{IJ}\}$ has two-form
in Plebanski sector (I$\pm$), (II$\pm$) or (deg), we say that $\{B_{ab}^{IJ}\}$,
relative to $\geosimp$, is also in Plebanski sector (I$\pm$), (II$\pm$) or (deg),
respectively.
\end{definition}
\noindent In fact, if  $\{B_{ab}^{IJ}\}$ is in a given Plebanski sector, this property is independent
of the choice of $\geosimp$, as we will now prove, so that the qualification
``relative to $\geosimp$'' is not necessary.
\begin{lemma}
If a given discrete Plebanski field $\{B_{ab}^{IJ}\}$  is in a given Plebanski sector
 ---  (I$\pm$), (II$\pm$) or (deg) --- relative to a given \ordname
$\geosimp$, then $\{B_{ab}^{IJ}\}$ is in the same Plebanski sector relative
to any \ordname.
\end{lemma}
{\startproof
Let a discrete Plebanski field $\{B_{ab}^{IJ}\}$  be given.
Let two ordered simplices $\geosimp$, $\geosimp'$ be given, and let
${}^{\geosimp}B_{\mu\nu}^{IJ}$, ${}^{\geosimp'}B_{\mu\nu}^{IJ}$
denote the corresponding two-forms.

There exists a unique element $G$ of the
inhomogeneous $GL(4)$ group, $IGL(4)$,
mapping the 5 vertices of $\geosimp$ into the 5 vertices of
$\geosimp'$, in order, so that $G$ maps $\geosimp$ into $\geosimp'$.
Furthermore, because the vertices of $\geosimp$ and $\geosimp'$ are each numbered
with positive orientation in the sense of definition \ref{ordsimpdef},
$G$ is in the \textit{proper}
inhomogeneous $GL(4)$ group, $IGL(4)^+$.
To construct, from $\{ B_{ab}^{IJ}\}$ and $\geosimp$, the two-form
${}^{\geosimp}B_{\mu\nu}^{IJ}$ on the manifold $M$, one only uses
the orientation of $M$ and the flat connection $\partial_a$; these are invariant under $IGL(4)^+$.
Therefore, for a given discrete Plebanski
 field  $\{B_{ab}^{IJ}\}$, the map from ordered 4-simplices $\geosimp$
 to two-forms ${}^\geosimp B_{\mu\nu}^{IJ}$ is $IGL(4)$ covariant.
 Thus one concludes
 \begin{displaymath}
G \cdot  {}^{\geosimp}B_{\mu\nu}^{IJ} = {}^{\geosimp'}B_{\mu\nu}^{IJ}.
 \end{displaymath}
But the action of $G$ in this equation is the action of a particular
diffeomorphism, and all diffeomorphisms preserves each
of the continuum Plebanski sectors (as is immediate from the diffeomorphism
covariance of the equation defining each one).
%
%
Thus, if either of
${}^{\geosimp}B_{\mu\nu}^{IJ}$,${}^{\geosimp'}B_{\mu\nu}^{IJ}$ is
in one of the Plebanski sectors, then both of them must be in the same Plebanski sector.
\finishproof}
Thus, the notion of discrete Plebanski field $\{B_{ab}^{IJ}\}$ being in a given
Plebanski sector is independent of the \ordname
used to define the continuum two-form.

\subsection{Plebanski sectors of the linear simplicity constraints}
\label{origsimp}

We here review the linear simplicity constraint used in the `new' spin-foam
models, and show that it
restricts the bivectors $B_{ab}$ to be in Plebanski sectors $(II\pm)$ or $(\text{deg})$.
The linear simplicity constraint imposes that
there exists an assignment of an $N^I_a$ to each $a$, such that
\begin{equation}
\label{discsimp}
C_{ab}^I := N_{aJ}\left(\dual B_{ab}\right)^{JI} \approx 0 \quad \forall b \neq a .
\end{equation}

Using this, we now define ``weak bivector geometry'' and ``bivector geometry''.
The latter is the same
as the definition in \cite{bdfgh2009}, \textit{except} that the $B_{ab}^{IJ}$ algebra elements
here are \textit{not} to be identified with the  $B_{ab}^{IJ}$ algebra elements in \cite{bdfgh2009},
but rather with their Hodge duals.  This is because we have chosen instead to be consistent
with the convention for  $B_{ab}^{IJ}$  used in \cite{epr2007, epr2007a, elpr2007, bhnr2004}.
%
%
%
\begin{definition}[Weak bivector geometry]
\label{weakbivdef}
A discrete Plebanski field $\{B_{ab}^{IJ}\}$ is additionally called
a \textit{weak bivector geometry} if
\begin{enumerate}
\item[(i.)] For each $a$ there exists $N_a^I \in \R^4$, such that
$N_{aI} (\dual B_{ab})^{IJ} = 0, \forall b \neq a$. (linear simplicity)
\item[(ii.)] For all distinct $a,b,c,d$, $\tr(B_{ab} [B_{ac},B_{ad}]) \neq 0$.
(tetrahedron non-degeneracy)
\end{enumerate}
\end{definition}

\begin{definition}[Bivector geometry]
\label{bivgeomdef}
A weak bivector geometry $\{B_{ab}^{IJ}\}$ is additionally
called a \textit{bivector geometry}
if it satisfies \textit{(full) non-degeneracy}:
For $i<j$, $i,j = 0,1,2,3$,
$B^{IJ}_{ij}$ form a basis of $\R^4 \otimes_{\mskew} \R^4$.
\end{definition}

The above definitions are intended to be applied to bivectors in the
\textit{4-simplex frame}.  The canonical variables, by contrast, are defined
in the tetrahedron frames, where we impose a gauge-fixed version of (\ref{discsimp})
in which $N_a^I$ is fixed to be $\scr{N}^I := (1,0,0,0)$, following \cite{elpr2007}.
In each tetrahedron frame, the simplicity constraint then becomes
\begin{equation}
\label{gfconstr}
C_{ab}^i := \half \epsilon^i{}_{jk} B_{ab}^{jk} \approx 0.
\end{equation}
In terms of $B_{ab}^\pm, \sut{\unitpm}_{ab}^\pm$ this is equivalent to
\begin{equation}
\label{linsimp_clcons}
B_{ab}^+ = B_{ab}^- \text{ and } \sut{\unitpm}_{ab}^+ = -\sut{\unitpm}_{ab}^-.
%
%
\end{equation}
The first of these equations is equivalent to what is called the ``diagonal simplicity
constraint''; as this is $SO(4)$ invariant, it is clear that it also follows from the
non-gauge-fixed condition (\ref{discsimp}).
Thus, although, in the original works, the constraint (\ref{discsimp}) was presented as
a reformulation of cross-simplicity only, in fact it also contains in it the diagonal
simplicity constraint.  This fact is also reflected in the quantum theory
in section \ref{EPRLsec}.
This is what allows us to omit diagonal simplicity as a separate condition
in definitions \ref{weakbivdef} and \ref{bivgeomdef}.

(\ref{linsimp_clcons})
implies that the solution space of (\ref{gfconstr}) can be parameterized by what we call
\textit{\cldataSUtwo}
$\{A_{ab}, \sut{\norm}_{ab}\}$:
\begin{equation}
\label{reducedvars}
B_{ab}^\pm = \half A_{ab} \text{ and } \sut{\unitpm}_{ab}^+ = -\sut{\unitpm}_{ab}^- = \sut{\norm}_{ab}.
\end{equation}
%
%
$A_{ab}$ and $\norm^i_{ab}$ have direct geometrical significance: $A_{ab}$ is the area of
triangle $\Delta_{ab}$, and $\norm^i_{ab}$ is the outward normal to $\Delta_{ab}$ in the $a$ frame.
Lastly, let
$B_{ab}^\phys(A_{ab}, \sut{n}_{ab}, X_{a}^\pm)$ denote
the corresponding bivectors $B_{ab}(\geosimp)$ in the \textit{4-simplex}
frame (\ref{Bsimp_param}):
\begin{equation}
\label{Bphys_def}
B_{ab}^\phys(A_{ab}, \sut{n}_{ab}, X_{a}^\pm) := B_{ab}(\simp)  = \half A_{ab}(-X_a^-\sut{\norm}_{ab}, X_{a}^+\sut{\norm}_{ab}).
%
%
\end{equation}
Note that all weak bivector geometries are of the form
$\{B^\phys_{ab}(A_{ab}, \sut{n}_{ab}, X^\pm_a)\}$
for some data set $\{A_{ab}, \sut{n}_{ab}, X^\pm_a\}$.
%
%

We next prove that a weak bivector geometry --- i.e., a discrete Plebanski field in which the linear simplicity constraint and tetrahedron non-degeneracy is imposed --- is in  Plebanski sector (II-), (II+) or (deg), and derive the conditions for each of these.
We do this by first proving a simpler theorem, quoting a
theorem from \cite{bdfgh2009}, and then proving the main result.
Remember here again that the
 $B_{ab}$ algebra elements used here are not the same as
 the $B_{ab}$ algebra elements used in \cite{bdfgh2009}, but
 are rather related by the Hodge dual.

We will use a canonical tetrad on $M \equiv \R^4$,
defined in the canonical chart as
$\mathring{e}_\alpha^I := \delta^I_\alpha$,
with associated metric $\mathring{g}_{\alpha\beta} = \mathring{e}^I_{\alpha} \mathring{e}_{\beta I}$ and
oriented volume form $\mathring{\epsilon}:= \mathring{e}^0 \wedge \mathring{e}^1 \wedge \mathring{e}^2
\wedge \mathring{e}^3$
given respectively by $\delta_{\alpha\beta}$ and the fully skew array with
$\mathring{\epsilon}_{0123} = 1$. Manifold indices will be raise and lowered with $\mathring{g}_{\alpha\beta}$

\begin{definition}[Geometrical bivectors]
\label{geombi}
Given \anordname in $M \equiv \R^4$, the associated
\textit{geometrical bivectors} $(B_{ab}^{\geom})^{IJ}$
are
\begin{equation}
(B_{ab}^{\geom})^{IJ}:= A(\Delta_{ab}) \frac{(N_a \wedge N_b)^{IJ}}{|N_a \wedge N_b|}
\end{equation}
%
%
where $A(\Delta_{ab})$ is the area of $\Delta_{ab}$ relative to $\mathring{g}_{\alpha\beta}$,
$N_a^\alpha$ is the outward pointing unit normal to the $a$th tetrahedron using
$\mathring{g}_{\alpha\beta}$, $N_a^I = \mathring{e}^I_\alpha N_a^{\alpha}$,
$(N_a \wedge N_b)^{IJ} := 2N_a^{[I} N_b^{J]}$, and
$|X^{IJ}|^2 := \half X^{IJ}X_{IJ}$.
\end{definition}
\noindent As pointed out in \cite{bdfgh2009}, for any \ordname $\geosimp$ in $M \equiv \R^4$,
$\{B_{ab}^{\geom}(\geosimp)\}$
form a bivector geometry.  Furthermore,
\begin{theorem}
\label{geomform}
For any \ordname $\geosimp$,  $\{B_{ab}^{\geom}(\geosimp)\}$
is in Plebanski sector (II+).
\end{theorem}
{\startproof
Let \anordname $\geosimp$ be given.
Let $N_a^\alpha$ denote the unit outward pointing normal to
the $a$th tetrahedron using $\mathring{g}_{\alpha\beta}$.
Let $\mathring{\epsilon}^{[ab]}_{\alpha\beta}$ denote the oriented
metric area form on $\Delta_{ab}$ and $\mathring{\epsilon}_{[ab]}^{\alpha\beta}$ its inverse.
From (\ref{triorient}),
\begin{displaymath}
\mathring{\epsilon}_{[ab]}^{\alpha\beta} = \lambda
(N_a)_\gamma (N_b)_\delta \mathring{\epsilon}^{\gamma\delta\alpha\beta}.
\end{displaymath}
for some positive function $\lambda$.
Let $\mathring{q}_{\alpha\beta}$ denote the pull-back of the
metric $\mathring{g}_{\alpha\beta}$ to $\Delta_{ab}$.  Then
\begin{eqnarray}
\nonumber
2 &=& \mathring{q}_{\alpha\gamma} \mathring{q}_{\beta\delta} \mathring{\epsilon}_{[ab]}^{\alpha\beta}
\mathring{\epsilon}_{[ab]}^{\gamma\delta}
= \mathring{g}_{\alpha\gamma} \mathring{q}_{\beta\delta} \mathring{\epsilon}_{[ab]}^{\alpha\beta}
\mathring{\epsilon}_{[ab]}^{\gamma\delta}\\
\nonumber
&=&  \lambda^{2} \mathring{g}_{\alpha\gamma} \mathring{g}_{\beta\delta}
(N_a)_\lambda (N_b)_\mu \mathring{\epsilon}^{\lambda\mu\alpha\beta}
(N_a)_\nu (N_b)_\sigma \mathring{\epsilon}^{\nu\sigma\gamma\delta} \\
\nonumber
&=& 4 \lambda^{2}
(N_a)_{[\lambda} (N_b)_{\mu]}
N_a^\lambda N_b^\mu
=  4 \lambda^{2}
(N_a)_{[I} (N_b)_{J]}
N_a^I N_b^J \\
\nonumber
&=&  \lambda^{2}
(N_a\wedge N_b)_{IJ}
(N_a\wedge N_b)^{IJ}\\
\nonumber
&=& 2 \lambda^{2}
|N_a\wedge N_b|^2
\end{eqnarray}
where, in the first line, we have used that $\mathring{\epsilon}_{[ab]}^{\alpha\beta}$
is tangent to $\Delta_{ab}$.
Thus
\begin{displaymath}
\lambda = |N_a\wedge N_b|^{-1} .
\end{displaymath}
So,
\begin{eqnarray*}
\int_{\Delta_{ab}} e^I \wedge e^J
&=& \int_{\Delta_{ab}} \mathring{\epsilon}^{[ab]} \mathring{\epsilon}_{[ab]}^{\alpha\beta} e^I_\alpha e^J_\beta
%
%
\\
&=&  |N_a\wedge N_b|^{-1} \int_{\Delta_{ab}} \mathring{\epsilon}^{[ab]}
(N_a)_\gamma (N_b)_\delta \mathring{\epsilon}^{\gamma\delta\alpha\beta} e_\alpha^I e_\beta^J \\
&=& |N_a\wedge N_b|^{-1} \int_{\Delta_{ab}} \mathring{\epsilon}^{[ab]}
(N_a)_K (N_b)_L \epsilon^{KLIJ} \\
%
%
&=& \half \epsilon^{IJ}{}_{KL} A(\Delta_{ab})
\frac{(N_a \wedge N_b)^{KL}}{|N_a\wedge N_b|}
\end{eqnarray*}
%
%
whence
\begin{displaymath}
B^\geom_{ab}(\geosimp)^{IJ} = \int_{\Delta_{ab}} \half \epsilon^{IJ}{}_{KL} e^K \wedge e^L
%
%
\end{displaymath}
with $\half \epsilon^{IJ}{}_{KL} e^K \wedge e^L$ constant.  This proves
that
%
%
$\{B_{ab}^\geom(\geosimp)\}$ is in Plebanski sector (II+).
\finishproof}

\noindent{}Let us review what can be called a partial version of
theorem 3 in \cite{bdfgh2009} (the `reconstruction theorem').  For the
following, we say that  a set of \cldataSUtwo
$\{A_{ab}, \sut{n}_{ab}\}$ is \textit{non-degenerate} if
for each $a$, every set of three vectors $\sut{n}_{ab}$ with $b\neq a$
is linearly independent, $\{A_{ab}, \sut{n}_{ab}\}$
satisfies closure if $\sum_{b\neq a} A_{ab} \sut{n}_{ab} = 0$,
and a set $\{X^\pm_a\} \subset SU(2)$ satisfies the \textit{orientation constraint}
if $X^\pm_a \sut{n}_{ab} = - X_b^\pm \sut{n}_{ba}$.
%
%
Furthermore,
\begin{definition}
Two sets of $SU(2)$ group elements $\{U^1_a\}, \{U^2_a\}$ are
called equivalent, $\{U^1_a\} \sim \{U^2_a\}$, if
$\exists \ggk \in SU(2)$ and a set of five signs $\epsilon_a$ such that
\begin{equation}
\label{eqrel}
U^2_a = \epsilon_a \ggk U^1_a.
\end{equation}
\end{definition}

\begin{theorem}[Partial version of the reconstruction theorem]
\label{reconth}
Let a set of non-degenerate \cldataSUtwo $\{A_{ab},\sut{n}_{ab}\}$ satisfying
closure be given, as well as
a set $\{X_a^\pm\} \subset SU(2)$, $a=1,\dots,5$, solving the
orientation constraint, such that
for each $a$, $X_a^- \not\sim X_a^+$.  Then there exists \anordname $\geosimp$ in $\R^4$, unique upto
inversion and translation,
such that
\begin{equation}
\label{reconpart}
B_{ab}^{\phys}(A_{ab},\sut{n}_{ab},X_a^\pm) = \mu B_{ab}^{\geom}(\geosimp)
\end{equation}
for some $\mu = \pm 1$, with $\mu$ independent of the ambiguity in $\geosimp$.
\end{theorem}
{\startproof
From the discussion in section 5.3.1 in \cite{bdfgh2009},
%
%
because
$\{A_{ab},\sut{n}_{ab}\}$ is a set of non-degenerate boundary data
and $X^-_a \not\sim X^+_a$ , $\{\dual B^\phys_{ab}(A_{ab}, \sut{n}_{ab}, X^\pm_a)\}$
satisfies non-degeneracy and hence is a bivector geometry in the sense of \cite{bdfgh2009},
and the result follows from theorem 3 in \cite{bdfgh2009}.  The fact that $\mu$ is independent of which
\ordname in $\R^4$ is used follows  from the invariance of $B^\geom_{ab}(\geosimp)$ under inversion and translation
of $\geosimp$.
\finishproof}

\begin{theorem}
\label{sectresult}
Suppose $\{A_{ab}, \sut{n}_{ab}\}$ is a set of non-degenerate \cldataSUtwo
satisfying closure and $\{X_a^\pm\}$ are such that orientation is satisfied.
\begin{enumerate}
\item[(i)]If $\{X^-_a\} \not\sim \{X^+_a\}$, then $\{B^{\phys}_{ab}(A_{ab}, \sut{n}_{ab}, X^\pm_a)\}$ is
either in Plebanski sector (II+) or (II-), according to whether the sign $\mu$
in theorem \ref{reconth} is $1$ or $-1$, respectively.
\item[(ii)]If $\{X^-_a\} \sim \{X^+_a\}$, then
$\{B^{\phys}_{ab}(A_{ab}, \sut{n}_{ab}, X^\pm_a)\}$ is
in the degenerate Plebanski sector.
\end{enumerate}
\end{theorem}
{\startproof

\noindent\textit{Proof of (i):}

This is immediate from theorems \ref{geomform} and \ref{reconth}.

\noindent\textit{Proof of (ii):}

As $\{X^-_a\} \sim \{X^+_a\}$, there exists signs $\epsilon_a$ and an
$SU(2)$ element $Y$ such that $X^+_a = \epsilon_a Y X^-_a$, so that
(\ref{Bphys_def}) becomes
\begin{displaymath}
B^\phys_{ab}(A_{ab}, \sut{n}_{ab}, X^\pm_a)
= \half A_{ab} (- X^-_a n_{ab}, (YX^-_a) n_{ab})
\end{displaymath}
where the sign $\epsilon_a$ could be dropped from the action
in the self-dual component because it is the adjoint action.
Let $H := (I, Y)$, let $\mathcal{N}^I = (1,0,0,0)$, and
let $N:= H \cdot \mathcal{N}$.  Then
\begin{eqnarray}
\nonumber
(\dual B^\phys_{ab})_{IJ} N^J =
(\dual B^\phys_{ab})_{IJ} (H \cdot \mathcal{N})^J
&=& (\dual B^\phys_{ab})_{KL} H^K{}_M H^L{}_J (H^{-1})^M{}_I \mathcal{N}^J \\
\nonumber
&=& (H^{-1} \triangleright \dual B^\phys_{ab})_{M0} (H^{-1})^M{}_I = 0
%
%
\end{eqnarray}
where (\ref{alginv}) has been used in the last line.
Let $\sigma$ be any ordered 4-simplex in $M$, and let $B_{\mu\nu}^{IJ}$
be the associated unique constant 2-form determined by $B_{ab}^\phys$.
Then we have
\begin{displaymath}
0 = \left(\int_{\Delta_{ab}} (\dual B)^{IJ}\right) N_J
= \int_{\Delta_{ab}} \left((\dual B)^{IJ} N_J\right)
\end{displaymath}
for all $a \neq b$. Because $(\dual B)^{IJ} N_J$ is constant, it follows that
$(\dual B)^{IJ}N_J \equiv 0$,  from which one can show
\begin{displaymath}
\eta^{\mu\nu\rho\sigma} \epsilon_{IJKL} B_{\mu\nu}^{IJ} B_{\rho\sigma}^{KL}
=
\eta^{\mu\nu\rho\sigma} \epsilon_{IJKL} (\dual B)_{\mu\nu}^{IJ}
(\dual B)_{\rho\sigma}^{KL} = 0
\end{displaymath}
so that $\{ B_{ab}^\phys \}$ is in the degenerate Plebanski sector.
%
%
\finishproof}

\section{Interpretation of the asymptotics of EPRL}
\label{qsect}

\subsection{Review of EPRL}
\label{EPRLsec}

Below, we give a brief review of the quantization leading to the EPRL vertex.
We do this both in order to clearly establish the meaning of the variables involved in
its definition, as well as to briefly remind the reader of the role of linear simplicity.

\subsubsection{Notation for $SU(2)$ and $Spin(4)$ structures.}

Given $g \in SU(2)$ and $\sut{x} \in \su(2)$, let $\rho_\genj(g), \rho_\genj(\sut{x})$
denote their representation on the spin $\genj$ carrying space $\spinsp_\genj$.
When it is clear from the context, the $\genj$ subscript will be dropped.
Let $\hat{\rotgen}^i:= i \rho(\sut{\tau}^i)$ denote the generators in the spin $\genj$ representation,
so that $[\hat{\rotgen}^i, \hat{\rotgen}^j] = i\epsilon^{ij}{}_k \hat{\rotgen}^k$.
%
%

Let $\epsilon : \spinsp_\genj \times \spinsp_\genj \rightarrow \C$ denote the standard skew-symmetric bilinear epsilon inner
product on $\spinsp_\genj$ \cite{rovelli2004, bdfgh2009}, satisfying
$\epsilon(\rho(g) \psi, \rho(g) \phi) = \epsilon(\psi,\phi)$ (and defined using the alternating spinor $\epsilon_{AB}$ when $\spinsp_\genj$
is realized as the symmetrized tensor product of $2k$ fundamental representations), and
let $\langle \cdot , \cdot \rangle$ denote the hermitian inner product on $\spinsp_\genj$, the
spin for these inner products being inferred from the arguments.
These inner products are related by the
antilinear structure map $J: \spinsp_\genj \rightarrow \spinsp_\genj$:
\begin{displaymath}
\langle \psi, \phi \rangle = \epsilon(J \psi, \phi).
\end{displaymath}
$J$ satisfies $J \rho_\genj(g) = \rho_\genj(g) J$, which implies $\hat{\rotgen}^i J = - J \hat{\rotgen}^i$.

Let $\spinsp_{\genjpm^-,\genjpm^+} = \spinsp_{\genjpm^-} \otimes \spinsp_{\genjpm^+}$ denote
the carrying space for the spin $(\genjpm^-,\genjpm^+)$ representation of $Spin(4) \equiv SU(2) \times SU(2)$,
and $\rho_{\genjpm^-,\genjpm^+}(X^-,X^+) := \rho_{\genjpm^-}(X^-)\otimes\rho_{\genjpm^+}(X^+)$ the representation of
$(X^-,X^+) \in Spin(4)$ thereon.  Again, when it is
clear from the context, the $\genjpm^-,\genjpm^+$ subscript will be dropped.
Let $\hat{J}^i_- := i\rho(\sut{\tau}^i) \otimes I_{\genjpm^+}$ and $\hat{J}^i_+ := i I_{\genjpm^-} \otimes \rho(\sut{\tau}^i)$
denote the anti-self-dual and self-dual generators
respectively, and let $\hat{\rotgen}^i := \hat{J}^i_- + \hat{J}^i_+$
denote the generators of spatial rotations on $\spinsp_{j^-, j^+}$.
%
%
Define the bilinear form $\epsilon : \spinsp_{\genjpm^+,\genjpm^-} \times \spinsp_{\genjpm^+,\genjpm^-} \rightarrow \C$ by
\begin{displaymath}
\epsilon(\psi^+ \otimes \psi^-, \phi^+ \otimes \phi^-) :=
\epsilon(\psi^+, \phi^+) \epsilon(\psi^-, \phi^-) .
\end{displaymath}
Finally, let $\iota_{\rotk}^{\genjpm^-, \genjpm^+}$ denote the intertwining map from
$\spinsp_{\rotk}$ to $\spinsp_{\genjpm^-} \otimes \spinsp_{\genjpm^+}$, unique upto scaling, with
scaling fixed by the requirement that it be isometric in the
Hilbert space inner products.
 %
 %

\subsubsection{Canonical data and phase space}

In the general boundary formulation of quantum mechanics \cite{rovelli2004},
to the boundary of any 4-dimensional region one associates a \textit{phase space}, which is then quantized to obtain
the \textit{boundary Hilbert space} of the theory formulated in that region \cite{rovelli2004}.
In the present case, the region is the 4-simplex $\simp$.
The boundary data is trivially constructed from the data introduced
in section \ref{discvarsect} ---
one has the algebra elements $B_{ab}$ and the related $J_{ab}$ in the frame
of each tetrahedron $a$, and for each pair of tetrahedra $a,b$ one constructs
a parallel transport map $G_{ab}$ from the frame $b$ the frame $a$.
These are related to the variables $G_a$ introduced in section
\ref{discvarsect} by $G_{ab} = (G_a)^{-1} G_b$.

These boundary data are assembled into a classical phase space which may be identified with
the cotangent bundle over any choice of five independent parallel transport maps $G_{ab} = (X_{ab}^+,X_{ab}^-)$,
$\Gamma = T^*(Spin(4)^5) = T^*((SU(2) \times SU(2))^5)$. Without loss of generality, we
choose these to be $G_{ab} = (X_{ab}^+,X_{ab}^-)$
with $a<b$.
For $a<b$, $J_{ab}=(J_{ab}^-,J_{ab}^+)$ and $J_{ba}=(J_{ba}^-,J_{ba}^+)$ respectively
generate right and left translations on $G_{ab}$.
%
%
For any $a\neq b$, we furthermore have the generators of
internal spatial rotations, $\rotgen^i_{ab} := \left(J^-_{ab}\right)^i + \left(J^+_{ab}\right)^i$.
Note that when linear simplicity is satisfied (\ref{gfconstr}), in terms of the
\cldataSUtwo we have
\begin{equation}
\label{rotgendata}
L^i_{ab} = \frac{1}{\kappa \gamma} A_{ab} n_{ab}^i .
\end{equation}
%
%

\subsubsection{Kinematical quantization and the vertex.}

The boundary Hilbert space of states $\Hil_{\partial\simp}^{Spin(4)}$ is the $L^2$ space over the five
$G_{ab}=(X_{ab}^-, X_{ab}^+) \in Spin(4)$ with $a<b$.
The momenta operators $(\hat{J}_{ab}^\pm)^i$ and $(\hat{J}_{ba}^\pm)^i$ act
respectively by $i$ times the right and left invariant vector fields,
associated to $\sut{\tau}^i \in \su(2)$,
on the elements $X_{ab}^\pm$.
A \textit{projected spin-network state} (see \cite{livine2002, alexandrov2007}) in
$\Hil_{\partial \simp}^{Spin(4)}$
is labeled by a choice of four spins $\genjpm_{ab}^\pm, \rotk_{ab}, \rotk_{ba}$ and
two states $\psi_{ab} \in V_{\rotk_{ab}}, \psi_{ba} \in V_{\rotk_{ba}}$ per triangle:
%
%
\begin{equation}
\label{projsn}
\Psi_{\{\genjpm_{ab}^\pm, \rotk_{ab}, \psi_{ab}\}}(X_{ab}^-, X_{ab}^+)
:= \prod_{a<b} \epsilon( \iota_{\rotk_{ab}}^{\genjpm_{ab}^-, \genjpm_{ab}^+}\psi_{ab},
\rho(X_{ab}^-, X_{ab}^+) \iota_{\rotk_{ab}}^{\genjpm_{ab}^-, \genjpm_{ab}^+}\psi_{ba}) .
\end{equation}
This is an eigenstate of $(\hat{J}_{ab}^\pm)^2=(\hat{J}_{ba}^\pm)^2$,
$\hat{\rotgen}_{ab}^2$ and $\hat{\rotgen}_{ba}^2$
with eigenvalues
$\genjpm_{ab}^\pm(\genjpm_{ab}^\pm + 1)$, $\rotk_{ab}(\rotk_{ab} + 1)$
and $\rotk_{ba}(\rotk_{ba} + 1)$, respectively, where
$\hat{\rotgen}_{ab}^2 = \sum_i(\hat{\rotgen}_{ab}^i)^2$.
The projected spin network states form an (overcomplete) basis
of $\Hil_{\partial \simp}^{Spin(4)}$.

To impose the linear simplicity constraint in quantum theory,
one takes the sum of the
squares of the constraints (\ref{gfconstr}) for each $ab$ to form a master constraint
\cite{thiemann2003, dt2004, dt2004a}:
\begin{equation}
\label{master}
\hat{M}_{ab} := \widehat{\sum_{i=1}^3 \left(C^i_{ab}\right)^2} .
\end{equation}
The ordering is determined by the stringent
condition that solutions exist.
The projected spin-networks (\ref{projsn}) are eigenstates of the resulting operator
$\hat{M}_{ab}$ with eigenvalue $M_{ab}$ given by
\begin{displaymath}
M_{ab} = \lambda\left[\left(1 - \frac{1}{\gamma^2}\right) \rotk_{ab}^2
+ \frac{2}{\gamma}\left(\frac{1}{\gamma} - 1\right)(\genjpm^+_{ab})^2
+ \frac{2}{\gamma}\left(\frac{1}{\gamma} + 1\right)(\genjpm^-_{ab})^2\right]
\end{displaymath}
where $\lambda$ is an unimportant positive constant.
%
%
From the constraints $\hat{M}_{ab} \Psi = 0$,
%
%
one derives
\begin{equation}
\label{finalsimp}
\rotk_{ab} = \frac{2 \genjpm^-_{ab}}{|1-\gamma|} = \frac{2 \genjpm^+_{ab}}{|1+\gamma|}
= \rotk_{ba}
\end{equation}
for all $a \neq b$.
The projected spin networks with labels satisfying quantum simplicity (\ref{finalsimp})
are  thus  parameterized
by a choice of one spin $\rotk_{ab}$ and two states $\psi_{ab}, \psi_{ba} \in V_{\rotk_{ab}}$ per triangle
--- exactly the parameters specifying a (generalized)
\textit{$SU(2)$ spin-network state} of LQG:
\begin{equation}
\label{spinnet}
\Psi_{\{\rotk_{ab},\psi_{ab}\}}(X_{ab}) := \prod_{a < b} \epsilon( \psi_{ab}, \rho(X_{ab}) \psi_{ba} ) \in \Hil^{LQG}_{\partial\simp}
\equiv L^2(SU(2)^5)
\end{equation}
The condition that $j^\pm_{ab} = \half |1 \pm \gamma| k_{ab}$ be
half-integer imposes a restriction on the spins $k_{ab}$;
let $\spinset$ be the set of allowable values of $\rotk_{ab}$, and let $\Hil_{\partial \simp}^{\gamma}$ be the span
of the $SU(2)$ spin-networks (\ref{spinnet}) with $\{\rotk_{ab}\} \subset \spinset$.
One has an isomorphism
$\iota: \Hil^{\gamma}_{\partial \simp} \rightarrow \Hil^{Spin(4)}_{\partial \simp}$
between $\Hil_{\partial \simp}^{\gamma}$
and the solution space to the master constraints in $\Hil^{Spin(4)}_{\partial \simp}$,
$\iota: \Psi_{\{\rotk_{ab}, \psi_{ab}\}} \mapsto \Psi_{\{\simpj_{ab}^\pm, \rotk_{ab}, \psi_{ab}\}}$,
where here, and throughout the rest of the paper, we set
$\simpj^\pm := \half|1\pm \gamma|\rotk$.\footnote{\label{gbfoot}The
solution space to the master constraints also satisfies the \textit{Gupta-Bleuler} criterion \cite{gupta1950, bleuler1950}
for the quantization $\hat{C}^i_{ab}$ of the \textit{original linear} simplicity
constraints $C^i_{ab}$ (\ref{gfconstr}): for all
$\Psi, \Psi' \in \Hil_{\partial \simp}^{\gamma}$
$\langle \iota \Psi, \hat{C}^i_{ab} \iota \Psi' \rangle = 0$
 \cite{dr2009}.
}
%
%

The EPRL vertex for a given LQG boundary state $\Psi_{\partial \simp}^{LQG} \in \Hil_{\partial \simp}^{\gamma} \subset \Hil_{\partial \simp}^{LQG}$
is then
\begin{equation}
\label{eprl}
A_v(\Psi^{LQG}_{\partial \simp})
= \int_{\rm Spin(4)^5} \prod_a \dif X^-_a \dif X^+_a
(\iota \Psi^{LQG}_{\partial \simp})(X^-_{ab}, X^+_{ab}).
\end{equation}

\subsection{Boundary coherent states and integral expressions}

\begin{definition}[Coherent state]
Given a unit 3-vector $n$ and a spin $\genj$, let
$|\Gamma(n)\rangle_\genj \in \spinsp_\genj$ denote the unit norm
state determined by the equation
$n \cdot \hat{L} | \Gamma(n) \rangle_\genj = j | \Gamma(n) \rangle_\genj$,
with phase ambiguity fixed arbitrarily for each $n$ and $\genj$.
For each $\theta$, define
$|n,\theta \rangle_\genj := e^{i\theta} | \Gamma(n) \rangle_\genj$.
These are the coherent states. The $\theta$ argument represents a phase ambiguity
that will usually be suppressed.
\end{definition}

\begin{definition}[\Qdata boundary data]
We call an assignment of one spin $k_{ab} \in \spinset$ and two unit
3-vectors $n_{ab}^i, n_{ba}^i$ per triangle $(ab)$ in $\simp$ a set of
\textit{\qdata boundary data}.
\end{definition}

\begin{definition}[Boundary state corresponding to a set of \qdata boundary data]
Given a set of \qdata  boundary data and a choice of phase $\theta$,
one defines a corresponding \textit{state} in the
SU(2) boundary Hilbert space of the simplex,
\begin{equation}
\label{genstate}
\Psi_{\{\rotk_{ab},n_{ab}\}, \theta} := \Psi_{\{\rotk_{ab}, \psi_{ab}\}}
\qquad \text{with} \qquad
|\psi_{ab}\rangle := |n_{ab}, \theta_{ab} \rangle_{\rotk_{ab}}
\end{equation}
where the $\theta_{ab}$ are any phases summing to $\theta$ modulo $2\pi$. The phase $\theta$ will usually
be omitted from the notation.
%
%
\end{definition}

In order to derive the asymptotics of the vertex, \cite{bdfgh2009}
first cast the vertex in appropriate integral
form, separately for the cases $\gamma < 1$ and $\gamma > 1$:
\begin{displaymath}
A_v(\Psi_{\{\rotk_{ab}, n_{ab}\}})
= \int \left(\prod_a \dif X^-_a \dif X^+_a\right) \exp(S_{\gamma < 1})
\qquad \text{for }\gamma < 1
\end{displaymath}
and
\begin{displaymath}
A_v(\Psi_{\{\rotk_{ab}, n_{ab}\}})
=
\int \left(\prod_a \dif X^+_a \dif X^-_a\right) \left(\prod_{ab} (-1)^{\simpj^-_{ab}} (2\simpj^+_{ab} + 1) \dif m_{ab}\right) \exp(S_{\gamma > 1})
\qquad \text{for }\gamma > 1
\end{displaymath}
where $\dif m_{ab}$ is the measure on the metric 2-sphere normalized to unit volume and
where $S_{\gamma < 1}$ and $S_{\gamma > 1}$ are ``actions'' \cite{bdfgh2009}.
These actions are generally complex.
As in \cite{bdfgh2009}, we are interested only in critical points whose contributions are not exponentially suppressed, and for this reason
define ``critical point'' to mean points where the action is stationary and its real part is maximal and \textit{non-negative}.
%
%
%
The critical point equations for $\gamma < 1$ are
\begin{eqnarray}
\label{orientcrit}
X^\pm_a \sut{n}_{ab} &=& - X_b^\pm\sut{n}_{ba} \\
 \label{closurecrit}
 \sum_{c \neq a} k_{ac}\sut{n}_{ac} &=& 0
\end{eqnarray}
for all $a,b$.  The critical point equations for $\gamma < 1$ are
again (\ref{orientcrit}) and (\ref{closurecrit}) \textit{plus}
equations determining $m_{ab}$ in terms of $n_{ab}$
%
%
Thus, the non-trivial critical point equations are always (\ref{orientcrit})
and (\ref{closurecrit}), allowing both cases to be treated in a unified way.

\subsection{Interpretation of the asymptotics and critical points}

Before interpreting the critical points in terms of Plebanski sectors, we make clear
the meaning of the data $\{k_{ab}, \sut{n}_{ab}, X^\pm_a\}$ in terms
of classical discrete geometry.
The data $\{k_{ab}, \sut{n}_{ab}\}$ label the coherent boundary state
$\Psi_{\{\rotk_{ab}, n_{ab}\}} \in \Hil^{LQG}_{\partial \simp}$, which,
in the definition of the vertex, is mapped by $\iota$ into
an $Spin(4)$ boundary state in $\Hil^{Spin(4)}_{\partial \simp}$.
By construction, $\iota \Psi_{\{\rotk_{ab},n_{ab}\}}$ satisfies linear simplicity
($\hat{M}_{ab} \iota \Psi = 0$).
Combined with
\begin{displaymath}
\langle \iota \Psi_{\{\rotk_{ab}, n_{ab}\}} | \hat{L}_{ab}^i
| \iota \Psi_{\{\rotk_{ab}, n_{ab}\}} \rangle
= k n_{ab}^i
\end{displaymath}
and equation (\ref{rotgendata}),
this leads to the conclusion that $\iota \Psi_{\{\rotk_{ab},n_{ab}\}}$
\textit{is a quantum state
approximating a $Spin(4)$ classical boundary state satisfying linear simplicity with
\cldataSUtwo $A_{ab}=A(\rotk_{ab}):= \kappa \gamma \rotk_{ab}$ and $\sut{n}_{ab}$.}\footnote{By looking
instead at the operator $\hat{L}^2$, one alternatively concludes
$A_{ab} = \tilde{A}(k_{ab}):= \kappa \gamma \sqrt{k_{ab}(k_{ab}+1)}$.  These two possibilities
for relating $A_{ab}$ and $\rotk_{ab}$ are equivalent in the semiclassical limit,
which is what concerns us here.}
Lastly, as \cite{bdfgh2009} do, we identify the group variables $X^\pm_a$ in
the definition of the vertex (\ref{eprl}) with the discrete connection introduced in section \ref{clframe}.
This identification is consistent
with the relation between the covariant and canonical transport variables presented
in section \ref{EPRLsec}.

We say that $\{\rotk_{ab}, \sut{n}_{ab}\}$ is \textit{non-degenerate}
or \textit{satisfies closure} iff $\{A(\rotk_{ab}), \sut{n}_{ab}\}$
is non-degenerate or satisfied closure, respectively.

\begin{definition}[Regge-like boundary data]
Let a non-degenerate \qdata boundary data set $\{\rotk_{ab}, \sut{n}_{ab}\}$
satisfying closure be given. Then for each tetrahedron $a$, there exists a
geometrical tetrahedron in $\R^3$, unique upto translations, such that each of the
four quantities $\{A(\rotk_{ab})\}_{b\neq a}$ is equal to the area of one of the triangular
faces, and each of the four vectors $\{n^i_{ab}\}_{b\neq a}$ is equal to the outward
pointing normal of the corresponding triangular face.  If these five geometrical
tetrahedra can be glued together consistently to form a 4-simplex, we say
that the boundary data is \textit{Regge-like}.
\end{definition}

If the data $\{\rotk_{ab}, \sut{n}_{ab}\}$ is Regge-like,  in particular this means that, for each pair of tetrahedra $a,b$, the triangle $ab$ in $a$ is congruent to the triangle $ba$ in $b$.
It follows that, for each pair of tetrahedra, there exists a unique $SU(2)$ element
$g_{ab}$ such that
(1.) the adjoint action of $g_{ab}$ on $\R^3$ maps the triangle $ab$ into the
triangle $ba$, and
(2.) $g_{ab} n_{ba} = - n_{ab}$, where $g_{ab}$ acts via the adjoint action.
It follows that $g_{ab} = g_{ba}^{-1}$.

\begin{definition}[Regge state]
If a \qdata boundary data set $\{\rotk_{ab}, \sut{n}_{ab}\}$ is Regge-like, then the phase ambiguity in
the state (\ref{genstate}) can be uniquely resolved \cite{bdfgh2009} by requiring that the phase of
the coherent states be chosen such that
\begin{displaymath}
g_{ab} |n_{ba} \rangle_{\rotk_{ab}} = J|n_{ab}\rangle_{\rotk_{ab}}
\end{displaymath}
%
%
The resulting state $\Psi_{\{k_{ab}, n_{ab}\}}$
is called the \textit{Regge state} corresponding to $\{k_{ab}, n_{ab}\}$, and we denote it
by $\Psi^{\Regge}_{\{k_{ab},n_{ab}\}}$.
\end{definition}

%
%
%

We are now ready to quote the EPRL
asymptotics from \cite{bdfgh2009}.
The statement of the asymptotics uses the fact that
the boundary geometry of a 4-simplex is sufficient to determine
the geometry of the 4-simplex itself \cite{bdfgh2009, connelly1993} and hence, in particular, the dihedral angles
$\Theta_{ab}$ between adjacent tetrahedra ---  if $N_a$ and $N_b$ denote
the outward pointing normals to the $a$th and $b$th tetrahedra, respectively, $\Theta_{ab}$ is
defined to be the unique angle in $[0,\pi]$ such that $N_a \cdot N_b = \cos \Theta_{ab}$.
For the following, we also need the notion of a \textit{vector geometry}:
A set of boundary data $\{k_{ab}, n_{ab}\}$ is called a \textit{vector geometry}
if it satisfies closure and there exists $\{h_a\} \subset SO(3)$ such that
$(h_a \cdot n_{ab})^i =  - (h_b \cdot n_{ba})^i$ for all $a \neq b$.
The notion of asymptotic here is the same as that in \cite{bdfgh2009}.

%
%
\begin{theorem}[EPRL asymptotics]
\label{asym_thm}
Let non-degenerate \qdata boundary data $\mathcal{B}=\{\rotk_{ab}, n_{ab}\}$ satisfying closure be given.
\begin{enumerate}
\item
If $\mathcal{B}$ is Regge-like, then in the limit $\lambda \rightarrow \infty$,
\begin{eqnarray}
\nonumber
\dummy \hspace{-0.5cm} A_v(\Psi^{\Regge}_{\{\lambda \rotk_{ab},n_{ab}\}}) \hspace{-2mm}&\sim&\hspace{-2mm} \left(\frac{2\pi}{\lambda}\right)^{12}\left[N_{+-}^\gamma \exp\left(i \sum_{a<b} A(\lambda \rotk_{ab}) \Theta_{ab}\right)
+
N_{+-}^\gamma \exp\left(- i \sum_{a<b} A(\lambda \rotk_{ab}) \Theta_{ab}\right)
\right.\\
\label{Regge_asym}
&&
\dummy \hspace{-2mm}
\left.+
N_{++}^\gamma \exp\left( \frac{i}{\gamma}  \sum_{a<b} A(\lambda\rotk_{ab}) \Theta_{ab}\right)
+
N_{--}^\gamma \exp\left( - \frac{i}{\gamma} \sum_{a<b} A(\lambda\rotk_{ab}) \Theta_{ab}\right)\right]
\end{eqnarray}
where $N_{+-}^\gamma, N_{++}^\gamma, N_{--}^\gamma$ are the Hessian factors given in \cite{bdfgh2009}.
\item
If $\mathcal{B}$ is not Regge-like, but forms a vector geometry,
 then in the limit $\lambda \rightarrow \infty$,
\begin{equation}
\label{vec_asym}
A_v(\Psi^{\Regge}_{\{\lambda \rotk_{ab},n_{ab}\}}) \sim \left(\frac{2\pi}{\lambda}\right)^{12} N
\end{equation}
where $N$ is as defined in \cite{bdfgh2009}.
\item
If $\mathcal{B}$ is not a vector geometry, then
$A_v(\Psi_{\{\lambda \rotk_{ab},n_{ab}\},\theta})$ decays exponentially with large $\lambda$ for any
$\theta$.
\end{enumerate}
\end{theorem}

\noindent\textit{Classification of the critical points according to Plebanski sector.}

We now come to the interpretation, in terms of Plebanski sectors,
of the critical points giving rise to the different terms in theorem \ref{asym_thm}:
\begin{itemize}
\item
At the critical points giving rise to the first two terms of (\ref{Regge_asym}), from \cite{bdfgh2009},
the data  $\{A(k_{ab}), n_{ab}, X^\pm_a\}$ satisfy $\{X^+_a\} \not\sim \{X^-_a\}$, and at the first term, $\mu = +1$,
while at the second term $\mu = -1$.   \textit{Therefore, by theorem \ref{sectresult},
the first two terms of (\ref{Regge_asym})
%
%
correspond to bivectors in Plebanski sectors
(II+) and (II-), respectively.}
\item
At the critical points giving rise to the rest of the non-exponentially suppressed
terms in theorem \ref{asym_thm}, from \cite{bdfgh2009}, the data  $\{A(k_{ab}), n_{ab}, X^\pm_a\}$
satisfy $\{X^+_a\} \sim \{X^-_a\}$. \textit{Therefore, by theorem \ref{sectresult},
the rest of the non-exponentially suppressed terms in the asymptotics ---
namely, terms 3 and 4 of (\ref{Regge_asym}), and (\ref{vec_asym}) ---
correspond to the degenerate Plebanski sector.}
\end{itemize}
The above statements constitute the principal conclusions of the present work.
%
%

\section{Conclusions}

In the foregoing work, we have clarified what it means for the discrete classical
data involved in the semiclassical interpretation of spin-foams to be in different
Plebanski sectors.  We then proved that
the simplicity constraint used in both EPRL and FK --- the linear simplicity constraint ---
restricts to Plebanski sectors (II+), (II-) and the degenerate sector,
mixing these three sectors.  Finally, after reviewing the asymptotics of the
 EPRL vertex, we have identified the Plebanksi sector of the data associated to each term in
the asymptotics.  This allowed us to see that the presence of terms other
than the desired $e^{i S_{\Regge}}$ term is directly due to the mixing of these
three Plebanksi sectors by linear simplicity. Although these
conclusions have been drawn for the Euclidean signature,
we expect similar arguments to hold in the Lorentzian case.

In the literature until now, when an interpretation of the different terms is given,
it is a different one. In the paper \cite{mp2011}, the viewpoint is mentioned that the presence of terms in the asymptotics with actions differing
only by a sign
are to be interpreted  as a \textit{sum over orientations
of the 4-simplex}.  This is based on an interpretation, first given in \cite{bdfgh2009} itself,
%
%
that the $\mu$ parameter in the reconstruction theorem (theorem \ref{reconth}) is
to be interpreted as measuring the orientation of the 4-simplex.
Although an interesting proposal, we believe this is not the natural interpretation:
For, what is relevant in distinguishing these critical points is the 
\textit{value of the discrete Plebanski field $B^{IJ}_{\mu\nu}(\simp)$ in the 4-simplex frame}
(or equivalently, in any one of the tetrahedron frames).
%
%
That is, based on the presentation in \cite{bdfgh2009} and here, the distinction between critical points with actions of equal and opposite value lie in the \textit{dynamical variables themselves}, 
and not in the orientation of the 4-simplex as a manifold.  Rather, we
have argued that the interpretation of such critical points is that of elements of two distinct
Plebanski sectors that are being mixed in the EPRL model.

For the purpose of  semiclassical calculations with the spin-foam model,
it is important that all terms in the asymptotics other than $e^{i S_{\Regge}}$
be eliminated. The only proposal so far in the literature for this
is to eliminate the extra terms by selecting the boundary state to be
\textit{peaked on the group variables as well as the conjugate canonical bivectors}
\cite{bmp2010, abr2009, rovelli2005}.
It is clear why this works: As mentioned, the different critical points in the terms
of the asymptotics differ in the values of the discrete Plebanski field 
in a chosen frame.  For the purpose of talking about boundary
states it is most convenient to use a tetrahedron frame so that
the discrete field depends only on the canonical data
$G_{ab}$ and $J^{IJ}_{ab}$. 
No matter which tetrahedron
frame is used, the discrete field will depend on \textit{both} the the \textit{group elements} $G_{ab}$ \textit{and} the \textit{conjugate canonical bivectors}
$J_{ab}^{IJ}$,
so that by choosing a boundary state peaked on both $J_{ab}^{IJ}$ and $G_{ab}$,
one is able to select a single discrete Plebanski field, and hence a single Plebanski sector,
and thus in particular to select the single term $e^{iS_{\Regge}}$ in the asymptotics, if desired.
%
%
Although this works for a single simplex, because
the strategy is based on specifying a boundary state, it is not immediately clear if
this solution will work for simplicial complexes with interior tetrahedra.

The conclusions of the present work suggest
another possible solution: If one could
modify the vertex in such a way as to restrict to only Plebanski sector (II+)
--- something which is necessary anyway in order to unambiguously describe
general relativity in the usual sense --- then the asymptotics of the vertex should be simply $e^{i S_{\Regge}}$, as desired. Such an avenue might be interesting to persue.

\section*{Acknowledgements}
The author thanks Christopher Beetle,
Frank Hellmann,
and Alejandro Perez
for discussions.

\bibliography{englebib}{}
\bibliographystyle{ieeetr}
%

\end{document}